\documentclass[english]{article}
\usepackage[T1]{fontenc}
\usepackage[latin9]{inputenc}
\usepackage{listings}
\usepackage{geometry}
\usepackage{float}
\usepackage{amsmath}
\usepackage{graphicx}
\usepackage{esint}

\makeatletter

\floatstyle{ruled}
\newfloat{algorithm}{tbp}{loa}
\providecommand{\algorithmname}{Algorithm}
\floatname{algorithm}{\protect\algorithmname}

\usepackage{url}

\makeatother

\usepackage{babel}
\begin{document}

\title{PyCraters: A Python framework for crater function analysis}

\author{Scott A. Norris}
\maketitle
\begin{abstract}
We introduce a Python framework designed to automate the most common
tasks associated with the extraction and upscaling of the statistics
of single-impact crater functions to inform coefficients of continuum
equations describing surface morphology evolution. Designed with ease-of-use
in mind, the framework allows users to extract meaningful statistical
estimates with very short Python programs. Wrappers to interface with
specific simulation packages, routines for statistical extraction
of output, and fitting and differentiation libraries are all hidden
behind simple, high-level user-facing functions. In addition, the
framework is extensible, allowing advanced users to specify the collection
of specialized statistics or the creation of customized plots. The
framework is hosted on the BitBucket service under an open-source
license, with the aim of helping non-specialists easily extract preliminary
estimates of relevant crater function results associated with a particular
experimental system.
\end{abstract}

\section{Introduction}

Irradiation by energetic ions is a ubiquitous materials processing
technique, used widely in laboratories and industry for doping, cleaning,
and modification of materials surfaces. Under certain environmental
conditions, ion irradiation is observed to induce the spontaneous
formation of nanometer-scale structures such as ripples, dots, and
holes \cite{navez-etal-1962}. In some contexts, such as the gradual
ion-induced degradation of fusion reactor components, these structures
are an artifact to be avoided \cite{baldwin-doerner-NF-2008}, but
more recent observations of well-ordered, high-aspect ratio structures
\cite{facsko-etal-SCIENCE-1999} has led to the consideration of ion
irradiation as an inexpensive means of inducing such structures deliberately.
With sufficient understanding, this process could serve as the basis
of an inexpensive, high-throughput means of creating surfaces with
desired mechanical, optical, or electronic properties, ready for immediate
application across the existing installed base of ion beam facilities.

A major barrier to predictive understanding of the ion-induced nanostructuring
process has been a large number of competing physical mechanisms,
and an accompanying difficulty in estimating their relative magnitudes
in different parameter regimes. To the original modeling of energy
deposition and its relationship to sputter yield by Sigmund \cite{sigmund-PR-1969,sigmund-JMS-1973},
and subsequent discovery of an sputter erosion-driven instability
mechanism identified by Bradley and Harper \cite{bradley-harper-JVST-1988},
numerous additional mechanisms have since been added. It has since
been discovered that atoms displaced during the collision cascade,
but not sputtered from the surface, produce contributions to the evolution
equations that directly compete with the erosive mechanism. First
studied in 1D by Carter and Vishnyakov, and later in 2D dimensions
by Davidovitch et al. \cite{carter-vishnyakov-PRB-1996,davidovitch-etal-PRB-2007},
this \textquotedblright{}momentum transfer\textquotedblright{} or
\textquotedblleft{}mass redistribution\textquotedblright{} contribution
effectively doubles the number of unknown parameters in the problem,
as every erosive term has a redistributive counterpart, and the magnitudes
of each must in principle be estimated. Intensifying this problem
was the realization that strongly-ordered structures generating the
most excitement seemed to be due to the presence of multiple components
in the target \cite{ozaydin-ludwig-JPCM-2009,macko-etal-NanoTech-2010},
whether through contamination, intentional co-deposition, or the use
of a two-component target such as III-V compounds. To accurately model
such equations requires the introduction of a second field to track
concentrations \cite{shenoy-chan-chason-PRL-2007} which, although
it can overcome the barriers to ordered structures exhibited by one-component
models \cite{bradley-shipman-PRL-2010,norris-JAP-2013-phase-separation},
again doubles the number of unknown parameters present in the problem.

Recently, the \textquotedbl{}Crater Function\textquotedbl{} framework
has emerged as a means of rigorously connecting surface morphology
evolution over long spatial and temporal scales to the statistical
properties of single ion impacts \cite{norris-etal-2009-JPCM,norris-etal-NCOMM-2011,norris-etal-NIMB-2013}.
Given the ``Crater Function'' $\Delta h\left(x,y;\mathcal{S}\right)$
describing the average surface modification due to an ion impact (with
a parametric dependence on the surface properties via the argument
$\mathcal{S}$), the multi-scale analysis within the framework \cite{norris-etal-2009-JPCM}
produces contributions to partial differential equations governing
the surface morphology evolution. It can therefore be viewed as a
way of estimating many of the unknown parameters present in these
problems by means of atomistic simulation. Originally applied only
to pure materials and using only data from flat surfaces \cite{norris-etal-NCOMM-2011},
the framework has since been expanded to the case of binary materials
\cite{norris-etal-NIMB-2013}, and to enable incorporation of simulation
data from curved targets \cite{harrison-bradley-PRB-2014}. It has
thus matured to the point where it may be of value to the general
community as a parameter estimation tool. However, to use the framework
to this end, one needs the capability to (a) perform numerical simulations
of ion impacts over a variety of surface parameters, (b) extract the
necessary statistics from the output of each simulations, (c) fit
parameter-dependent statistics to various appropriate functional forms,
and (d) combine and report the results. 

This manuscript describes a Python library to provide all of the
above capabilities through a simple and user-friendly API. Access
to various simulation tools is provided via wrappers that automatically
create input files, run the solver, read the output, and save in a
common format. A customizable set of statistical analyses are then
run on the common-format output file, and saved for later use under
a unique parameter-dependent filename. A flexible loading mechanism
and general purpose fitting library make it easy to load statistics
as a function of arbitrary parameters, and then fit the resulting
data points to appropriate smooth nonlinear functions. Finally, functions
utilizing these capabilities are provided to perform all current mathematical
operations indicated by the literature to extract and plot PDE components.
Using this library, example codes that re-obtain existing results
within the literature are as short as 20 lines each, and can easily
be modified by end users to begin studying systems of their choosing.

\section{Theoretical Background}

\paragraph*{Crater Functions.}

If the dominant effects of the impact-induced collision cascade can
be assumed to take place near to the surface of the evolving film,
then the normal surface velocity of an ion-bombarded surface can be
represented by an integro-differential equation of the form \cite{aziz-MfM-2006},
\begin{equation}
v_{n}=\int I\left(\phi\left(\mathbf{x}^{\prime}\right)\right)\Delta h\left(\mathbf{x}-\mathbf{x}^{\prime};\mathcal{S}\left(\mathbf{x}^{\prime}\right)\right)\,\mathrm{d}\mathbf{x}^{\prime},\label{eq: crater-integral-equation}
\end{equation}
where $I=I_{0}\cos\left(\phi\right)$ is the projected ion flux depending
on the local angle of incidence $\phi\left(\mathbf{x}\right)$, $\Delta h$
is the ``crater function'' describing the average surface response
due to single ion impacts, and $\mathcal{S}$ describes an arbitrary
parametric dependence of the crater function on the local surface
shape. This form has advantages over more traditional treatments of
irradiation-induced morphology evolution. Instead of separate, simplified
models of the processes of ion-induced sputtering \cite{sigmund-JMS-1973,bradley-harper-JVST-1988}
and impact-induced momentum transfer \cite{carter-vishnyakov-PRB-1996,davidovitch-etal-PRB-2007}
-- both of which break down as the angle of incidence approaches grazing
-- the crater function $\Delta h$ naturally includes components due
to both sputtered atoms and redistributed atoms (thus unifying the
two approaches), and can in principle be obtained empirically (thus
avoiding inaccuracy at high angles of incidence).

\paragraph*{A Generic Framework.}

Exploiting the typical experimental observation of a separation of
spatial scales between the size of the impact (direct spatial dependence
of $\Delta h$) and the typical size of emergent structures (spatial
dependence of $\phi$ and $\mathcal{S}$), a multiple-scale analysis
was conducted in which the integral in Eq.~(\ref{eq: crater-integral-equation})
is expanded into an infinite series of terms involving the moments
of $\Delta h$ \cite{norris-etal-2009-JPCM}:
\begin{equation}
v_{n}=\left[I\tilde{M}^{\left(0\right)}\right]+\varepsilon\nabla_{S}\cdot\left[I\tilde{M}^{\left(1\right)}\right]+\frac{1}{2}\varepsilon^{2}\nabla_{S}\cdot\nabla_{S}\cdot\left[I\tilde{M}^{\left(2\right)}\right]+\dots\label{eq: norris-2009-1}
\end{equation}
where $I$ is the projected ion flux (a scalar), the $\nabla_{S}$
are co-ordinate free surface divergences, and the $\tilde{M}^{\left(i\right)}$
are ``effective'' moments of the crater function $\Delta h$ in
increasing tensor order \cite{norris-etal-2009-JPCM}. This result
provides intuition as to which parts of the crater function $\Delta h$
are most important to understand morphology evolution, and represents
a general solution for the multiple-scale expansion of Eq.~(\ref{eq: crater-integral-equation})
in the sense that it should apply for any parametric dependencies
of the crater function on the surface shape $\mathcal{S}$. (Note
that in this co-ordinate free form, the effective moments are really
combinations of the actual moments as described in Ref.~\cite{norris-etal-2009-JPCM};
however, in any linearization, they are equivalent).

\paragraph*{Example Applications.}

Equation~(\ref{eq: norris-2009-1}) is fully nonlinear and independent
of the specific form of the crater function. Therefore, to study surface
stability, one must first choose a form for the crater function $\Delta h$,
and then linearize the resulting specific instance of Equation~(\ref{eq: norris-2009-1})
about a flat surface. This process was first demonstrated in Ref.~\cite{norris-etal-NCOMM-2011},
where for simplicity and consistency with available simulation data,
the crater function
\begin{equation}
\Delta h=g\left(\mathbf{x}-\mathbf{x}^{\prime};\phi\right),\label{eq: simple-delta-h}
\end{equation}
was chosen, which depends parametrically only on the local angle of
incidence $\phi$. Inserting this expression into the general result
Eq.~(\ref{eq: norris-2009-1}), linearizing, and then adopting a
moving frame of reference to eliminate translational and advective
terms, one obtains to leading order the PDE
\begin{equation}
\frac{\partial h}{\partial t}=S_{X}\left(\theta\right)\frac{\partial^{2}h}{\partial x^{2}}+S_{Y}\left(\theta\right)\frac{\partial^{2}h}{\partial y^{2}}\label{eq: simple-PDE}
\end{equation}
 where the angle-dependent coefficients $S_{X}\left(\theta\right)$
and $S_{Y}\left(\theta\right)$ are related to the crater functions
via the expressions

\begin{equation}
\begin{aligned}S_{X}\left(\theta\right) & =\frac{\partial}{\partial\theta}\left[I_{0}\cos\left(\theta\right)M_{x}^{\left(1\right)}\left(\theta\right)\right]\\
S_{Y}\left(\theta\right) & =\left[I_{0}\cos\left(\theta\right)\cot\left(\theta\right)M_{x}^{\left(1\right)}\left(\theta\right)\right]
\end{aligned}
\label{eq: SXY-natcomm}
\end{equation}
where $M_{x}^{\left(1\right)}$ is the component of the (vector) first
moment in the projected direction of the ion beam. More recently,
a similar approach has been applied to an extended crater function
of the form \cite{harrison-bradley-PRB-2014}
\begin{equation}
\Delta h=g\left(\mathbf{x}-\mathbf{x}^{\prime};\phi,K_{11},K_{12},K_{22}\right)\label{eq: bradley-extended-crater}
\end{equation}
depending additionally on the surface curvatures $K_{ij}$ near the
point of impact. It was found that including this dependency within
the crater function reveals additional terms in the coefficient values,
which take the revised form
\begin{equation}
\begin{aligned}S_{X}\left(\theta\right) & =\frac{\partial}{\partial\theta}\left[I_{0}\cos\left(\theta\right)M_{x}^{\left(1\right)}\left(\theta\right)\right]+\frac{\partial}{\partial K_{11}}\left[I_{0}\cos\left(\theta\right)M^{\left(0\right)}\right]\\
S_{Y}\left(\theta\right) & =\left[I_{0}\cos\left(\theta\right)\cot\left(\theta\right)M_{x}^{\left(1\right)}\left(\theta\right)\right]+\frac{\partial}{\partial K_{22}}\left[I_{0}\cos\left(\theta\right)M^{\left(0\right)}\right]
\end{aligned}
.\label{eq: SXY-bradley}
\end{equation}

\paragraph*{Implications.}

The practical consequence of such results is that one can directly
connect atomistic simulations over small length- and time-scales to
continuum equations governing morphology evolution over much longer
scales. If the crater function $\Delta h$ and its moments $M^{\left(i\right)}$
can be identified as functions of the local surface configuration
(angle, curvature, etc.) by simulation (e.g. \cite{bringa-etal-PRB-2001})
or even experiment (e.g. \cite{costantini-etal-PRL-2001}), then the
expected continuum evolution of the system can be predicted via Eq.~(\ref{eq: simple-PDE})
with coefficients supplied by results such as Eqs.~(\ref{eq: SXY-natcomm})
or (\ref{eq: SXY-bradley}). Early applications of these ideas show
significant promise for predicting the outcome of experiments \cite{norris-etal-NCOMM-2011}
or determining the likely physical mechanisms driving experimental
observations \cite{norris-etal-NIMB-2013}. However, the steps required
to do so represent a non-trivial task in simulation and data analysis:
an effective procedure must address questions of 
\begin{description}
\item [{(1)}] creation or selection of a simulation tool to perform many
single-impact simulations
\item [{(2)}] obtaining statistically converged moments at individual parameter
combinations, 
\item [{(3)}] estimation of derivative values using data from adjacent
parameter combinations
\item [{(4)}] a smoothing mechanism to prevent uncertainties at step (2)
from being amplified in step (3)
\end{description}
An approach incorporating such steps was first demonstrated in Ref.~\cite{norris-etal-NCOMM-2011},
where molecular dynamics simulations using the PARCAS code \cite{nordlund-parcas}
were performed for irradiation of Si by 250 eV Ar+ at 5-degree increments
between $0^{\circ}$ and $90^{\circ}$. Smoothing was accomplished
by a weighted fitting of the simulation results to a truncated Fourier
series, and fitted values of $M_{x}^{\left(1\right)}\left(\theta\right)$
were then inserted into Eqs.~(\ref{eq: SXY-natcomm}). Analyzing
the resulting linear PDE of the form (\ref{eq: simple-PDE}) (with
additional terms describing ion-enhanced viscous flow), the most-unstable
wavelengths at each angle were compared to the wavelengths of experimentally-observe
structures, with reasonable agreement. In the process, the relative
sizes of the effects of erosion and redistribution were directly obtained
and compared, and the effects of redistribution were unexpectedly
found to be dominant for the chosen system.

\section{Goals and Outline of the Framework}

The previous section outlines the general features of the Crater Function
approach, including the potential promise for the general problem
of coefficient estimation, and also the technical hurdles associated
with its use. However, while the process of simulation, statistical
analysis, fitting, and differentiation is time-consuming, it is also
in principle mechanical, suggesting the utility of an open-source
library to centralize best practices and avoid repeated re-implementations.
Here we describe the primary goals of such a library, and a summary
of the structure of the resulting codebase.

\paragraph*{Motivation \#1: Accessibility.}

A principal motivation for the present work is to automate the process
described above in a generic and accessible way. As many of the procedures
as possible should be performed automatically, with reasonable default
strategies applied for users that do not wish to delve into the details
of atomistic simulation, internal statistical data structures, or
the optimal fitting functions for moment curves. Instead, a first-time
user should be able to spend most of their effort on deciding what
system and environmental parameters to study, after which the library
should take care of subsequent mechanical operations. In particular,
it should provide simple visualization routines for tasks expected
to be common, such as plotting the calculated angle-dependent coefficient
values.

\paragraph*{Motivation \#2: Extensibility.}

A second motivation is to accommodate the desires of users who wish
to move beyond the basic capabilities just described. For example,
whereas first-time users may wish to call a high-level function to
obtain a graph similar to one already published in the literature,
a more advanced user might like to work directly with the raw statistical
data, using the collected moments to extract particular quantities
of interest. Finally, a very advanced user or researcher within the
field may wish to customize the set of statistics to be gathered,
or even modify the methods used to calculate and fit those customized
statistical quantities. As much as possible, the goal should be to
accommodate users at each of these levels of detail, and allowing
each of them to use built-in utilities to simplify the calculation,
loading, fitting, and plotting of whatever quantities are of interest
to a given user.

\paragraph*{Motivation \#3: Portability.}

A final motivation is to enable the collection of statistical data
from as many sources as possible, with as little effort as possible.
When first demonstrated in Ref.~\cite{norris-etal-NCOMM-2011}, simulations
were performed by Molecular Dynamics (MD), using the PARCAS simulation
code \cite{nordlund-parcas}. However, in principle any MD code could
be used, such as the open-source LAAMPS code \cite{laamps-md-code}.
Furthermore, if the time required for MD simulation is a significant
obstacle, then the simpler, faster Binary Collision Approximation
(BCA) becomes an appealing approach. A variety of BCA codes exist,
including many descended from the widely-used SRIM code \cite{biersack-haggmark-NIM-1980-srim,ziegler-biersack-littmark-1985-SRIM,ziegler-SRIM-website}
such as TRIDYN \cite{moeller-eckstein-NIMB-1984-TRIDYN,moeller-tridyn-hzdr-website},
SDTRIMSP \cite{mutzke-etal-2011-sdtrimsp-5} and TRI3DST \cite{nietiadi-etal-PRB-2014}.
Importantly, these various codes have different and complementary
capabilities, and so practitioners may need to employ different codes
to answer different types of questions. To facilitate this, the library
should in principle be compatible with as many as possible.

\paragraph*{Framework Summary.}

In response to these motivations, we have therefore implemented our
library as a layered framework. All of the analysis routines are implemented
using an internal, standardized set of data structures for holding
the results of simulation output. These structures, themselves, are
then hidden behind a set of commonly-used high-level routines for
generating coefficient values, which may be easily used by practitioners
without reproducing any of the underlying work. Furthermore, the framework
is designed to be agnostic with respect to the simulation tool used
to obtain the impact data. Any solver capable of producing such data
can be ``wrapped'' within an input-output object whose job is simply
to write the input files required by that solver, run the simulation,
and read the resulting output files. Because input and output file
formats vary widely between codes, the specifics of each wrapper are
abstracted from the user. Instead, simulation parameters common to
all solvers are available in a simplified, standardized, high-level
user interface common to all wrappers, whereas features specific only
to certain solvers can be specified on a solver-by-solver basis. 

\begin{figure}
\centering{}\includegraphics[width=5in]{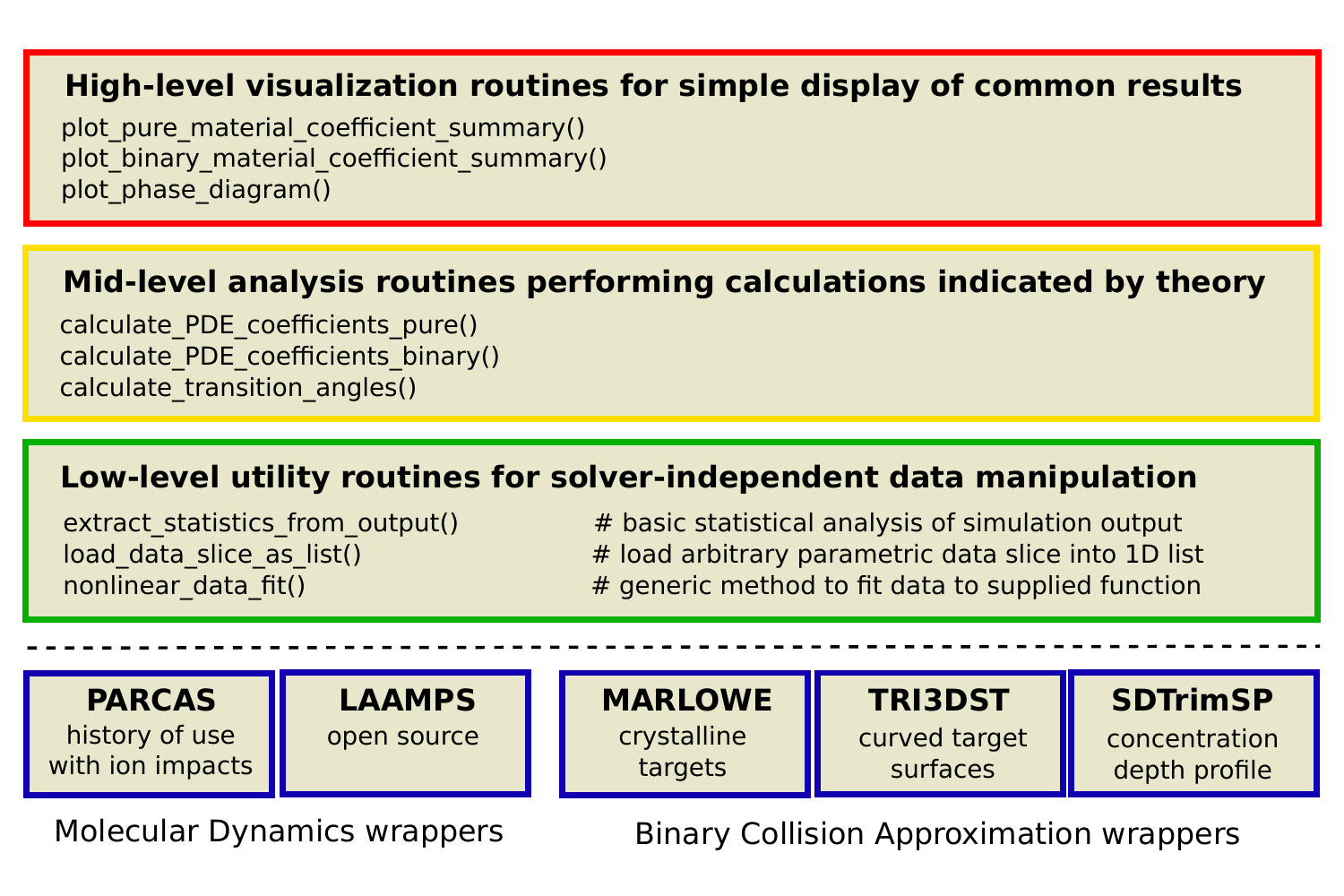}\caption{A schematic diagram of the organization of the PyCraters library.
(a) At the lowest level are solver-specific wrappers that abstract
the performance of individual simulations behind a standardized interface.
Requests for simulations are performed, and results obtained, in a
standard format. (b) Above this layer is a set of generic utilities
for the collection of statistics from simulation data, the storage
and loading of data in easy-to-use formats, and the fitting of data
points to smooth curves. (c) Next is a layer for performing the real
work associated with the crater function formalism -- the loading
of appropriate slices of data, fitting that data to appropriate curves,
and differentiating the resulting fits to obtain coefficients of linearized
equations. (d) Finally, a related set of visualization utilities plots
commonly-sought coarse-grained quantities. On top of this framework,
small end-user codes to perform basic surveys can easily be written
by non-specialists, while specialists can directly invoke the lower-level
routines to suit their particular needs.\label{fig: pycraters-schematic}}
\end{figure}

The overall approach is illustrated in Figure~\ref{fig: pycraters-schematic},
and consists of components at various levels of abstraction. From
lowest-level to highest-level, these may be summarized as consisting
of
\begin{enumerate}
\item Wrappers around individual Molecular Dynamics or Binary Collision
Approximation solvers.
\item Generic, but extensible analysis routines for the extraction of moments
and other statistics.
\item An customizable library for the fitting of these statistics to appropriate
smooth functional forms.
\item Routines for performing calculations needed to convert fitted functions
into PDE coefficients.
\item Plotting utilities to display reasonable summaries of various kinds
of data.
\end{enumerate}
It should be stressed, however, that from the user's perspective,
these capabilities are visible in the opposite order. For example,
a user desiring to plot the PDE coefficients for a given material
would simply call an associated plotting function. This function checks
to see if fits are available from a prior use of the script, and if
not, requests such fits from a lower-level function. That function
checks to see whether the statistical data are available, and if not,
requests them from a still lower-level function. Finally, that function
invokes a BCA or MD simulation tool wrapper, which performs all communication
with the solver needed to obtain the statistics.

We conclude this section by noting that our goal is not to completely
wrap the functionality of sophisticated tools such as LAAMPS and TRIDYN
-- rather, it is to provide a consistent interface to these tools
for users with the specific aim of using them to gather and apply
single-impact crater function statistics. Python -- a high-level scripting
language with very mature numerical capabilities -- is an ideal language
in which to implement such an interface. We note that although certain
kinds of parameter sweeping are built into some of the solvers we
have discussed, such capabilities effectively represent miniature
``scripting languages'' unique to each solver. By using a mature
language like Python to perform all scripting, the PyCraters library
allows parameter sweeping and statistical analysis to be done in a
uniform manner, regardless of the underlying tool used to perform
the simulations, and therefore without having to learn the details
of the scripting capabilities of each solver.

\section{Usage Examples}

We will now proceed to briefly illustrate several examples of the
framework in use. The focus of this section is not on any particular
result obtained herein (which are subject to revision as theoretical
approaches improve), nor on precisely documenting the codebase (which
is subject to change in future software versions). Rather, it is to
emphasize the general structure of the code as depicted in Figure~\ref{fig: pycraters-schematic},
and illustrate the kinds of problems that the framework has been designed
to investigate. In all cases, we will estimate PDE coefficients using
the simpler Eqs.~(\ref{eq: SXY-natcomm}), procedures for which have
been documented in the literature. A report on somewhat more complicated
procedures for using the revised Eqs.~(\ref{eq: SXY-bradley}) will
be the topic of future work.

\subsection{Preliminaries: Basic library loading and setup}

We begin with a very brief introduction to the code typically needed
at the beginning of a PyCraters script, shown in Algorithm~\ref{alg: basic-setup}.
It shows the loading of common mathematical libraries, as well as
a wrapper and set of helper routines from the PyCraters library. The
first code block simply loads all needed libraries, including the
relevant parts of PyCraters itself. The second code block reads the
executable location from the command line, and creates the two main
objects needed to interact with the library -- a solver wrapper and
a parameter holder. The latter is tasked with describing the simulation
environment, while the former abstracts each solver behind a uniform
interface. Note that in these two code blocks, the user chosen the
TRI3DST solver \cite{nietiadi-etal-PRB-2014}. If a different solver
were desired, only three lines of code would need to be changed. 

\begin{algorithm}
\begin{lstlisting}[basicstyle={\ttfamily},numbers=left]
# import necessary libraries
import sys
import numpy as np
import matplotlib.pyplot as plt 
import pycraters.wrappers.TRI3DST as wrap
import pycraters.helpers as help
import pycraters.IO as io

# build solver wrapper and parameter object
exec_location = sys.argv[1]
wrapper = wrap.TRI3DST_Wrapper(exec_location)
params  = wrap.TRI3DST_Parameters()
\end{lstlisting}
\caption{Common code needed to load the PyCraters libraries.\label{alg: basic-setup}}
\end{algorithm}

\subsection{Simplest example: Angle-dependence over one energy\label{sub: simple-example-Ne-C}}

We now present a very simple illustration of the framework, by using
it to obtain results of the kind reported in Ref~\cite{norris-etal-NCOMM-2011}
-- PDE coefficients associated with the low-energy irradiation of
a pure material. For narrative consistency with the next section,
we choose 100 eV $\mathrm{Ne}^{+}\to\mathrm{C}$. The code needed
for this example appears in Algorithm~\ref{alg: example-usage},
and should be assumed to be preceded by the code in Algorithm~\ref{alg: basic-setup}.
The first code block specifies the environmental parameters the user
wishes to consider, including the ion and target species, the ion
energy and incidence angle, and the number of impacts to perform (the
incidence angle is left blank). The second code block sweeps over
the incidence angle in 5-degree increments: for each angle, it updates
the parameter holder, and calls the solver wrapper for the specified
set of parameters. Finally, the third code block calls a routine that
plots a simple summary of the results.

\begin{figure}
\centering{}\includegraphics[width=6in]{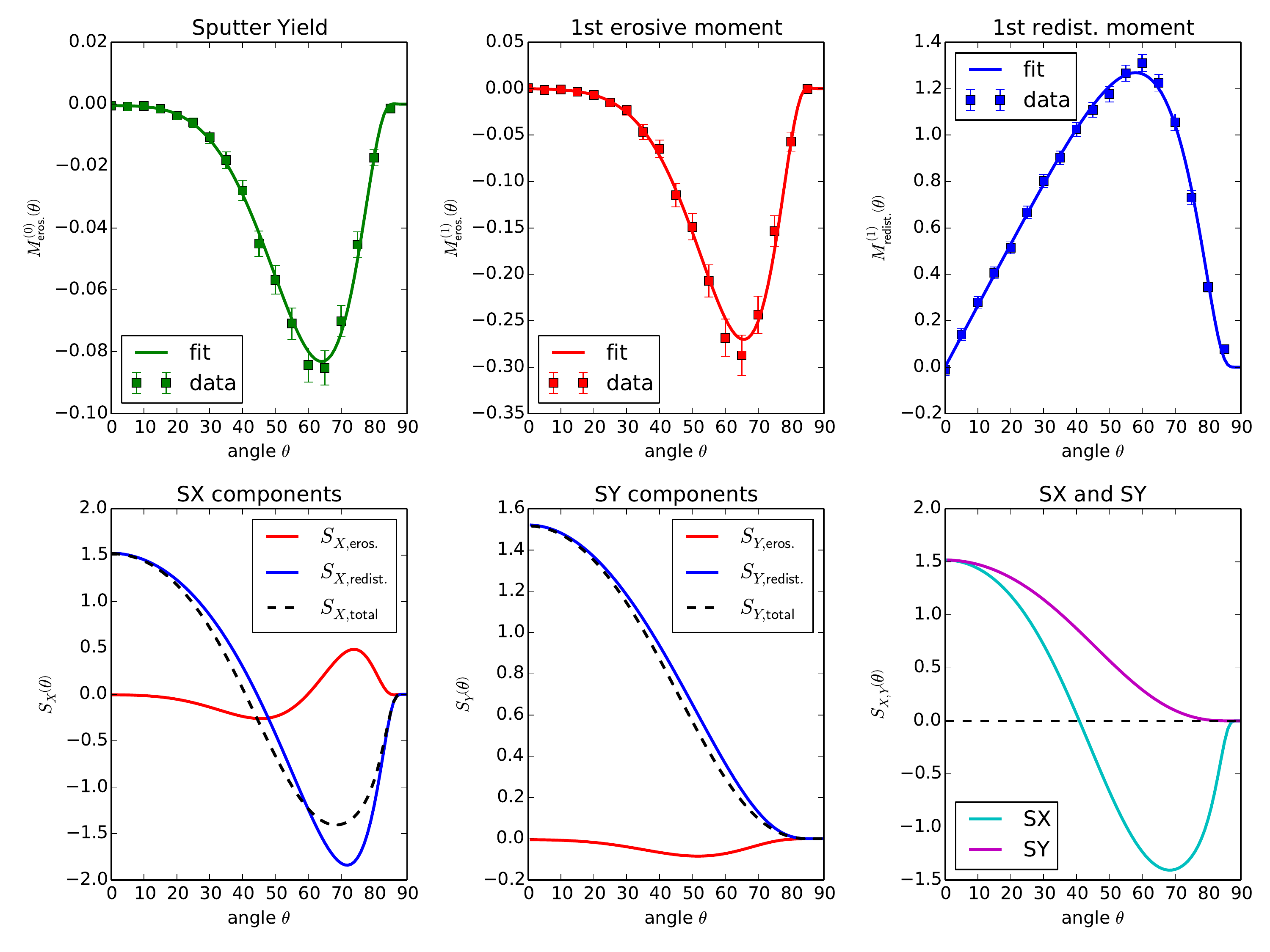}\caption{Output of the program listing in.\label{fig: example-output}}
\end{figure}

The bulk of the work performed by the PyCraters framework is hidden
behind just three function calls. The call to the \texttt{wrapper.go()}
routine performs all interaction with the underlying BCA solver, including
writing input files, calling the executable, reading output files,
extracting moments, and storing the results on the hard disk under
a unique file name constructed by the \texttt{Params()} object. Next,
the call to the \texttt{help.extract\_PDE\_coefficients()} routine
reads these files for all angles in the sweep, uses a self-contained
library to fit each of the angle-dependent moments to appropriate
functions of the incidence angle, performs differentiations needed
to construct the coefficients, and stores both fits and coefficients
under additional reconstructable filenames. Finally, the \texttt{help.plot\_coefficient\_summary()}
routine is a relatively simple visualization routine containing plots
likely to be of interest to a casual user. For the program listing
in Algorithm~\ref{alg: example-usage}, the output is exhibited in
Fig.~\ref{fig: example-output}.

\begin{algorithm}
\begin{lstlisting}[basicstyle={\ttfamily},numbers=left]
# set parameter values in high-level format 
params.target  = [["C", 1.0]]
params.beam    = "Ne" 
params.energy  = 100 
params.angle   = None 
params.impacts = 1000 

# perform simulations over a series of angles
angles = np.linspace(0,85,18) 
for aa in angles:
  params.angle = aa
  wrapper.go(params)

# extract statistics, perform fits, and generate plots 
fitdata = help.extract_PDE_coefficients(params, angles) 
results = help.plot_coefficient_summary(fitdata) 
plt.show()
\end{lstlisting}
\caption{An example program listing using the PyCraters Python framework.\label{alg: example-usage}}
\end{algorithm}

\subsection{Angle-Energy Phase Diagrams}

In this section we present a slightly more complicated example, demonstrating
the ease with which additional parameters may be swept to identify
trends in statistical behavior. Specifically, we observe that because
the stability of Eqn.~\ref{eq: simple-PDE} is determined by the
signs of the angle-dependent coefficients $S_{X}\left(\theta\right)$
and $S_{Y}\left(\theta\right)$, the sixth panel of Figure~\ref{fig: example-output}
describes the transition in expected behavior from flat surfaces (both
coefficients positive) to ripples oriented in the $x$-direction ($S_{X}<0$).
The points at which these curves cross the origin, and each other,
thus serve to divide domains of different expected behavior. Using
the PyCraters framework, it is only a matter of adding an extra \texttt{for()}
loop to repeat this calculation for a variety of ion energies, and
thereby obtain an angle-energy phase diagram. Furthermore, additional
sweeps over ion and target species allow the associated phase diagrams
for each ion/target combination to be compared, enabling the identification
of trends in stability with respect to ion and target atom mass. The
code used to perform these simulation is listed as Algorithm~\ref{alg: phase-diagram-listing}.
We note the great similarity to the code listed in Algorithm~\ref{alg: example-usage}
-- with the exception of three extra nested \texttt{for()} loops,
and the additional calls to functions like \texttt{help.find\_pattern\_transitions()}
(which calculates curve intersections between $\left\{ 0,\, S_{X}\left(\theta\right),\, S_{Y}\left(\theta\right)\right\} $)
and \texttt{help.plot\_energy\_angle\_phase\_diagram()} (which plots
the results) -- the majority of the code is similar. This reflects
the aims of the framework of providing high-level functionality in
easy-to-use functions, allowing the end user to focus on specifying
the range of parameters to be explored.

\begin{figure}
\begin{centering}
\includegraphics[width=3in]{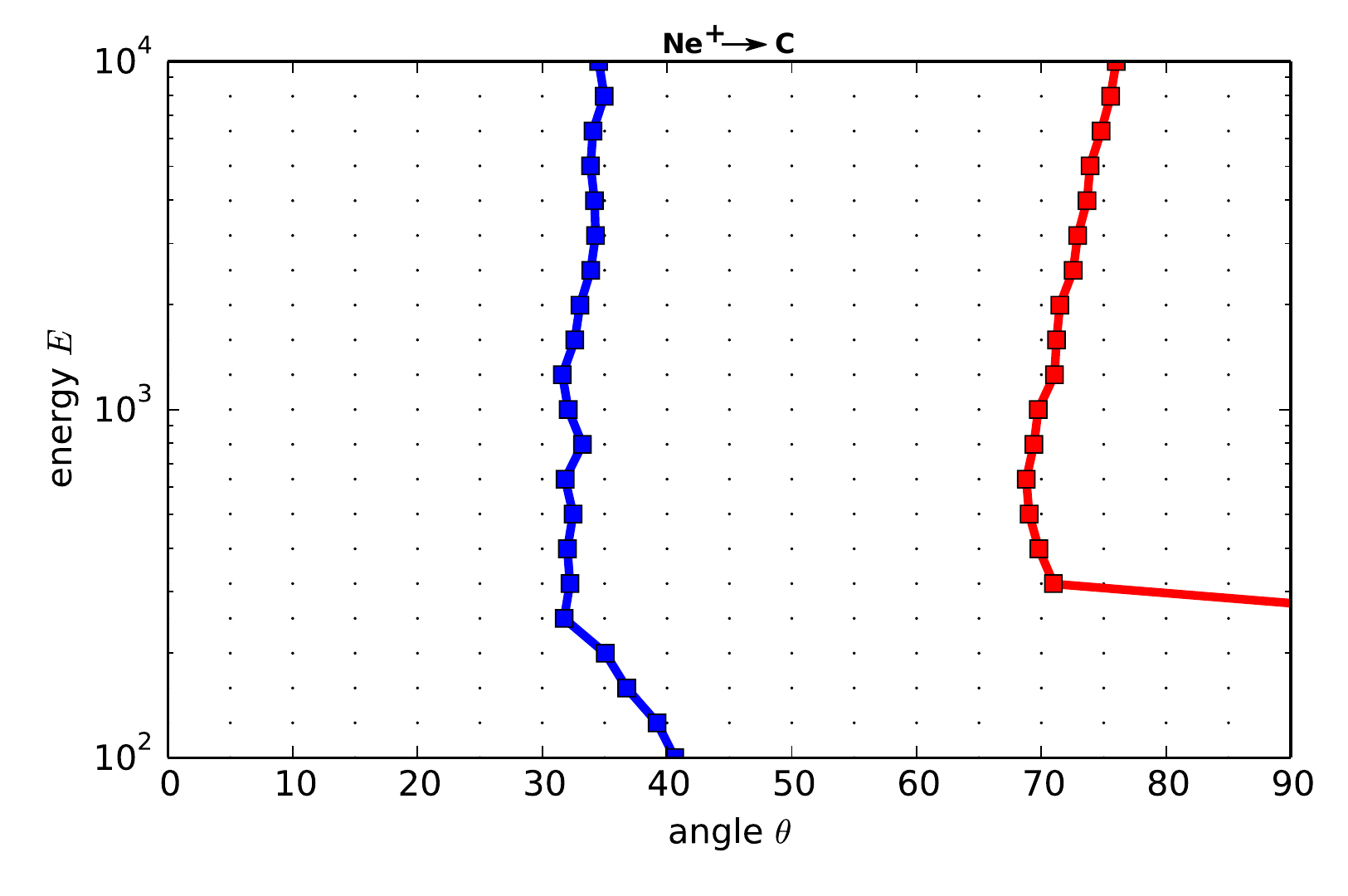}\includegraphics[width=3in]{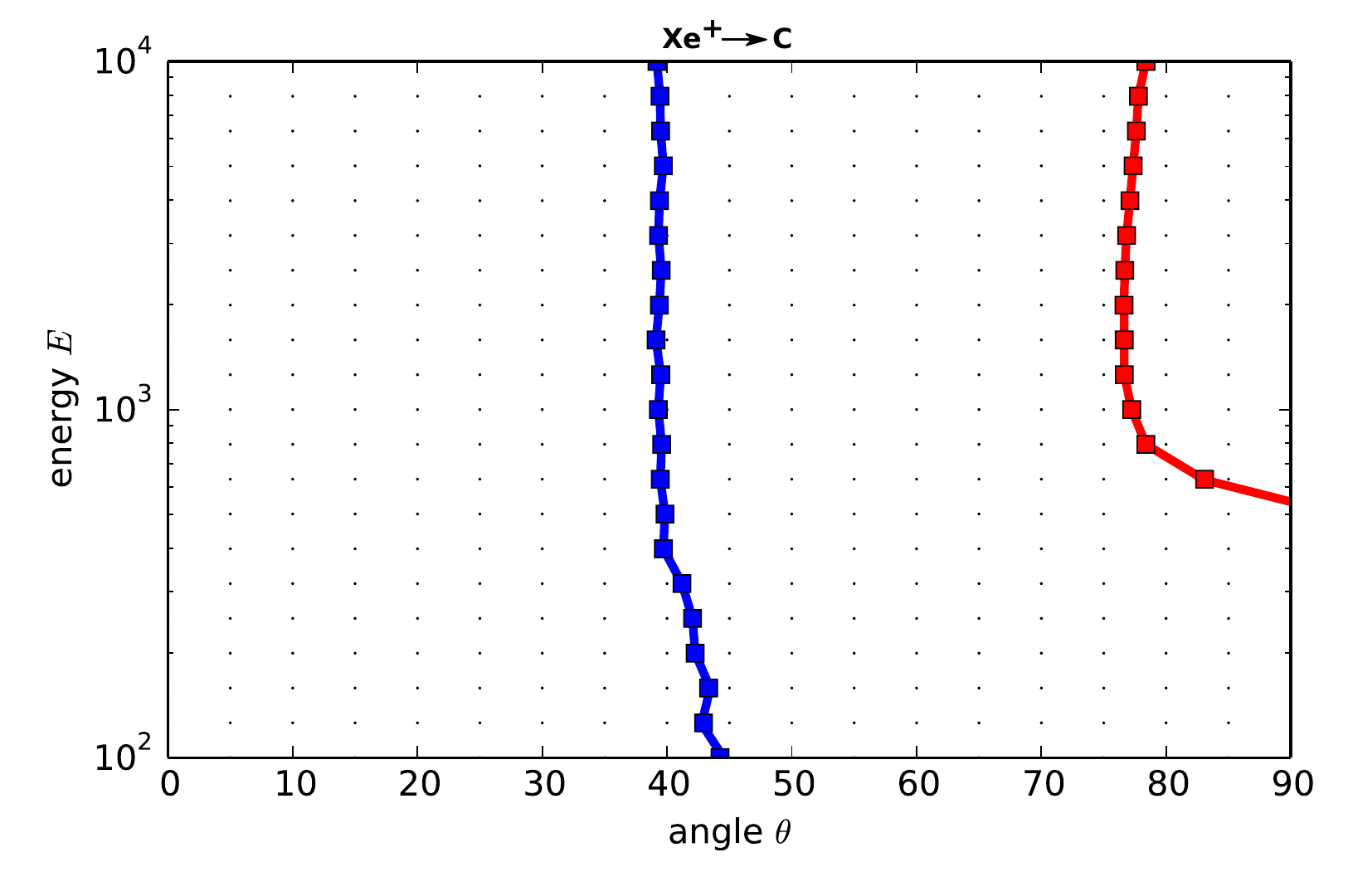}
\par\end{centering}

\begin{centering}
\includegraphics[width=3in]{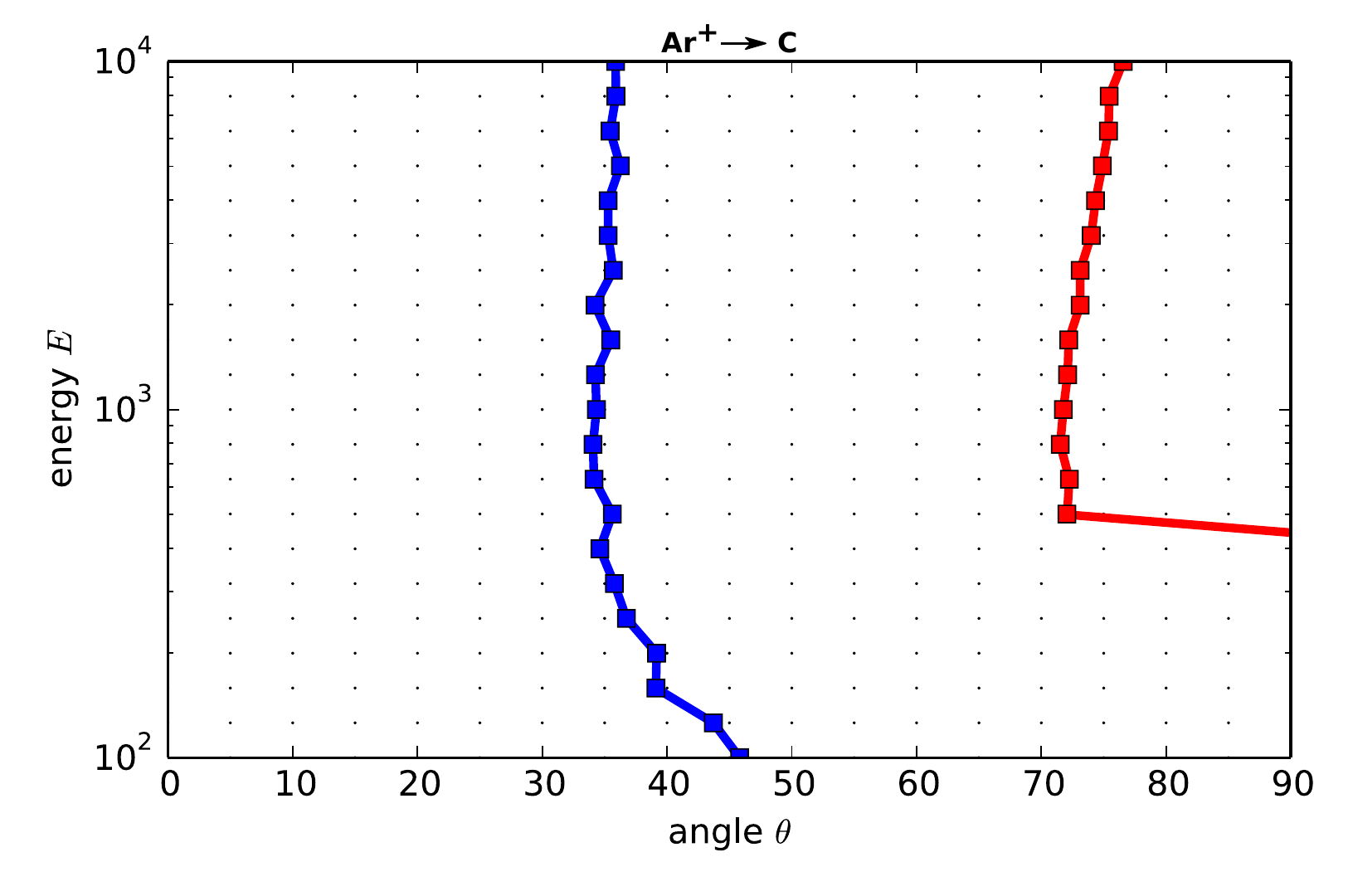}\includegraphics[width=3in]{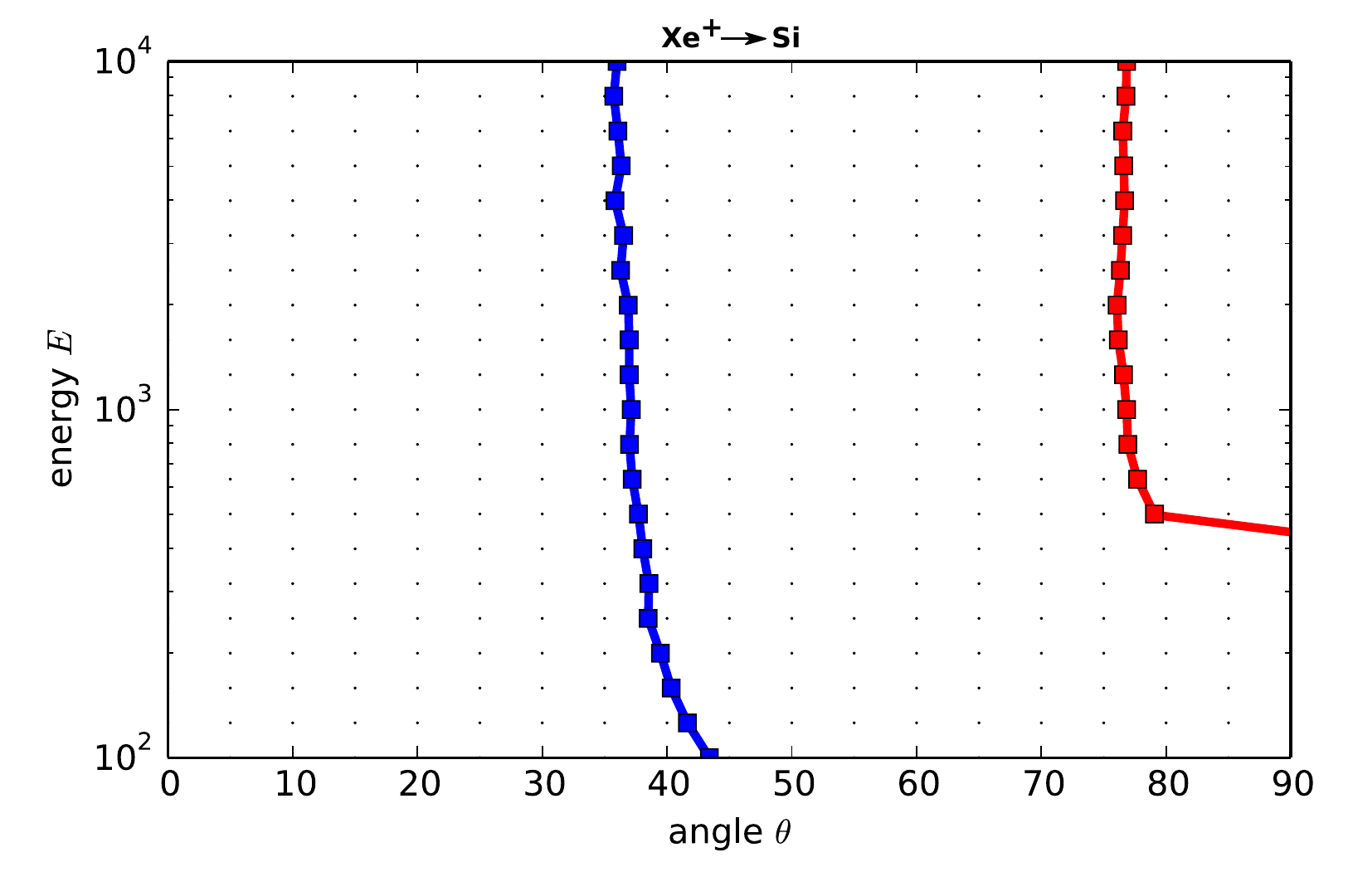}
\par\end{centering}

\begin{centering}
\includegraphics[width=3in]{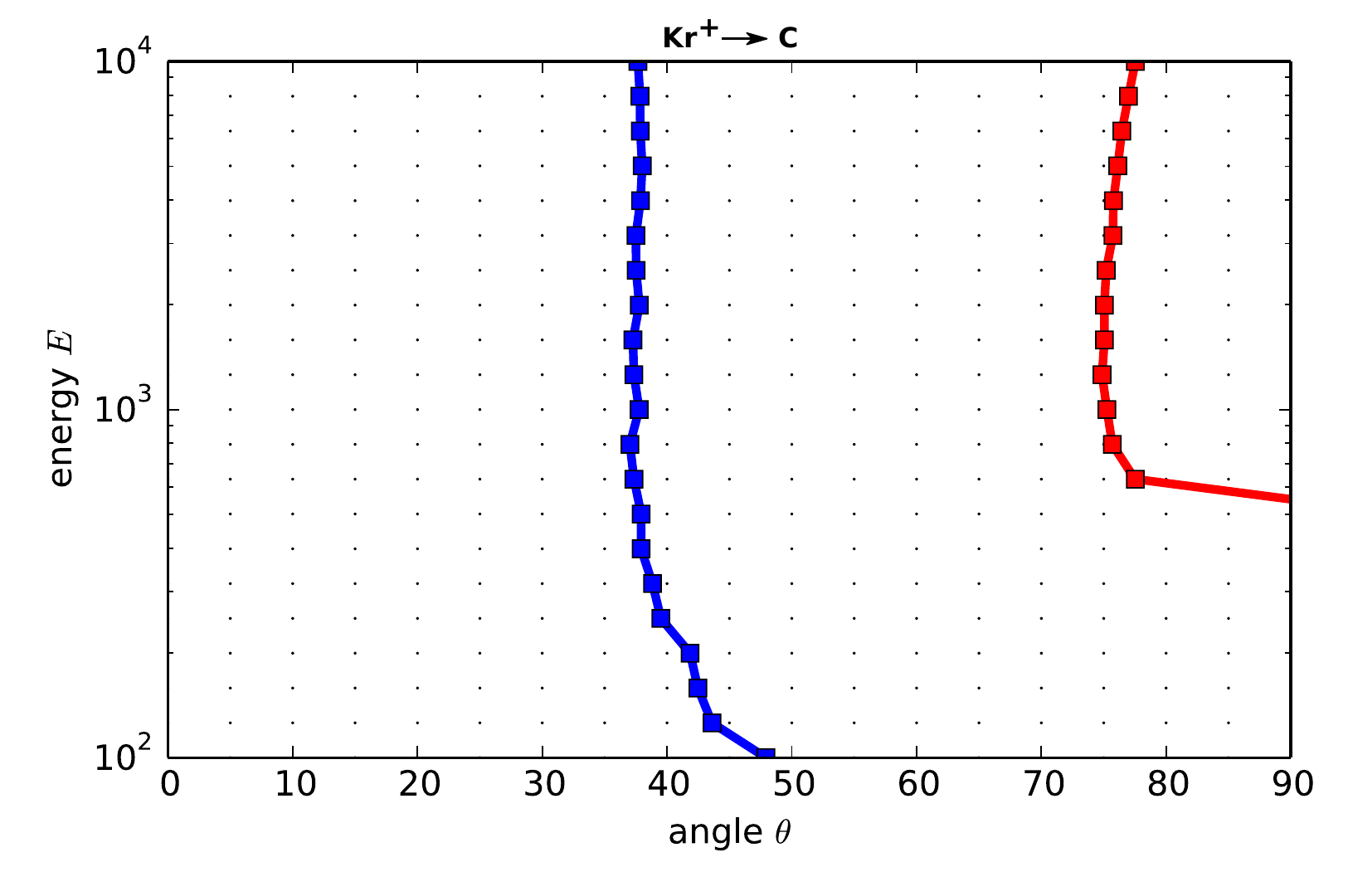}\includegraphics[width=3in]{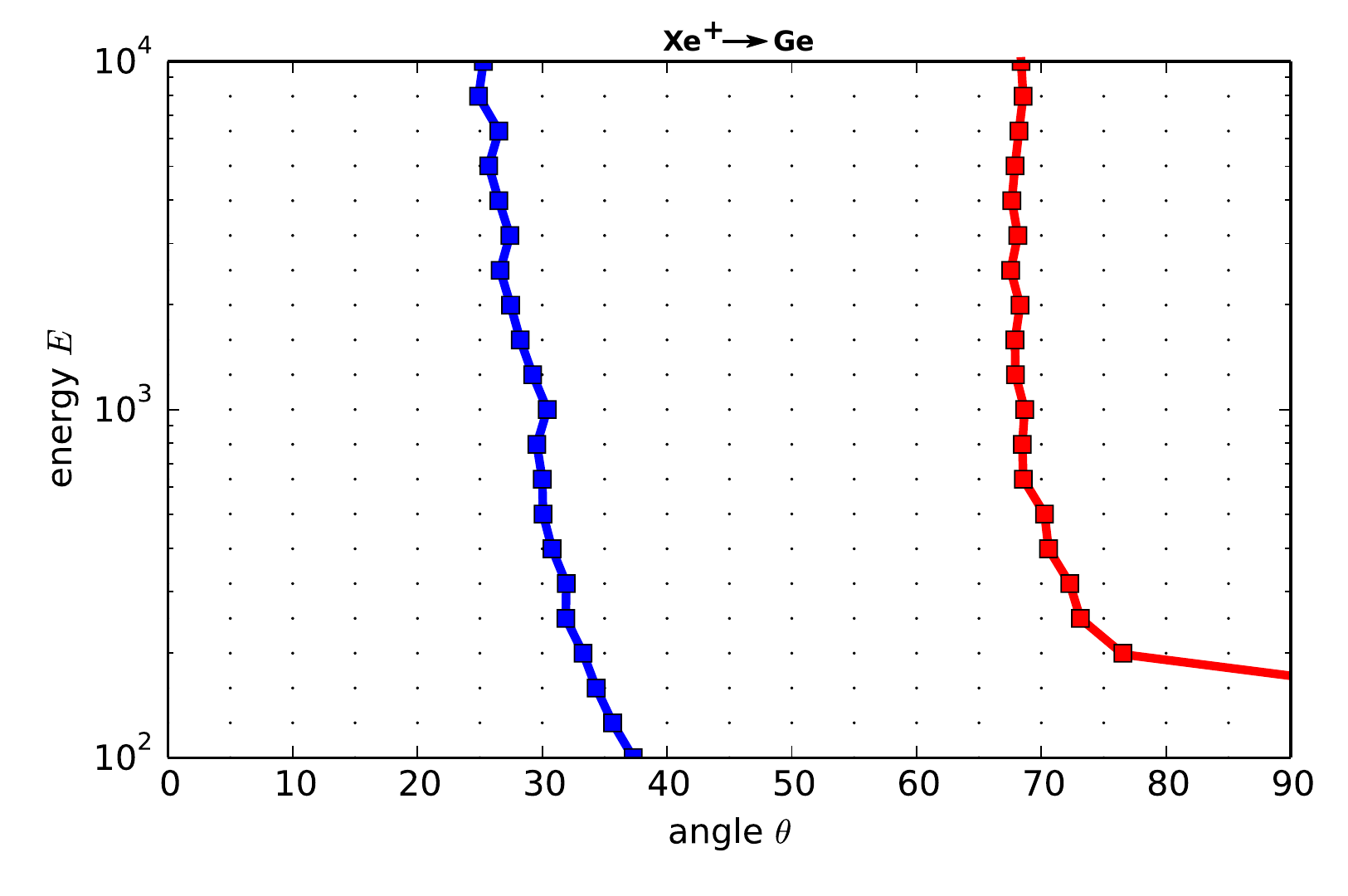}
\par\end{centering}

\centering{}\includegraphics[width=3in]{XeC-phase-diagram}\includegraphics[width=3in]{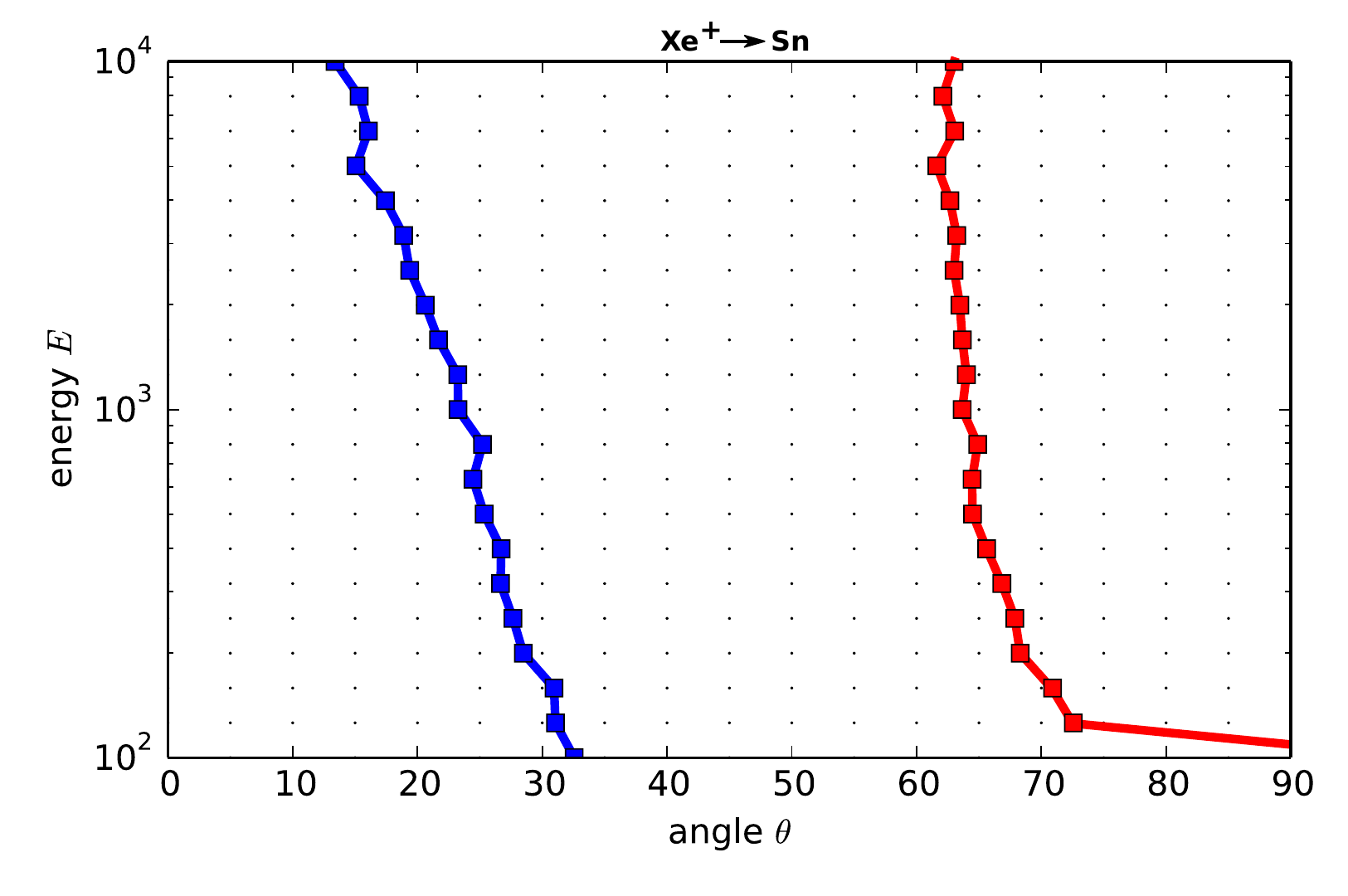}\caption{A collection of angle-energy phase diagrams generated using the program
listed in Algorithm~\ref{alg: phase-diagram-listing}. Left column
descending: increasing ion mass ($\left\{ \mathrm{Ne}^{+},\,\mathrm{Ar}^{+},\,\mathrm{Kr}^{+},\,\mathrm{Xe}^{+}\right\} \to\mathrm{C}$).
Right column descending: increasing target mass ($\mathrm{Xe}^{+}\to\left\{ \mathrm{C},\,\mathrm{Si},\,\mathrm{Ge},\,\mathrm{Sn}\right\} $).
In each figure, the region between the left edge and the blue line
indicates flat targets, the region between the blue and red lines
indicates parallel mode ripples, and the region between the red line
and the right edge indicates perpendicular mode ripples.\label{fig: phase-digram-collection}}
\end{figure}

A sampling of the respective graphs are shown in Figure~(\ref{fig: phase-digram-collection}).
In the first column, the ion mass is increased, and a corresponding
increase in the size of the region for stable, flat surfaces is observed,
as well as a decrease in the size of the region for perpendicular
mode ripples. Because the stable regions are induced by redistribution,
and the latter by erosion, this indicates an increasing relative strength
of the redistributive effect as the ion mass increases. By contrast,
in the second column, the target mass is increased, and the trend
is reversed -- the stable region shrinks, and the region of perpendicular
mode ripples grows. This indicates that generically, as the target
mass increases, the role of erosion grows relative to that of mass
redistribution. Interestingly, for most ion/mass combinations, the
effect of the ion energy seems to be small above around 500 eV. It
should be stressed, however, that these results are not presented
as a definitive prediction on behavior. They capture only the effect
of the collision cascade -- sputter erosion and mass redistribution
-- and do not capture phenomena such as stress implantation and relaxation
\cite{castro-cuerno-ASS-2012,norris-PRB-2012-viscoelastic-normal,norris-PRB-2012-linear-viscous,castro-etal-PRB-2012}.
This effect also contributes to the coefficients $S_{X}\left(\theta\right)$
and $S_{Y}\left(\theta\right)$, and its magnitude relative to collisional
effects is not yet known.

\begin{algorithm}
\begin{lstlisting}[basicstyle={\ttfamily},numbers=left]
# identify parameters to sweep over
targets  = [ [["C", 1.0]], [["Si", 1.0]], [["Ge", 1.0]], [["Sn", 1.0]]]
beams    = ["Ne", "Ar", "Kr", "Xe"]
energies = 10.**(np.linspace(2.0, 4.0, 21))
angles   = np.linspace(0,85,18) 
finedeg  = np.linspace(0, 90, 91)
impacts  = 1000

# perform the simulations and store the results
for tt in targets:
  for bb in beams:
    tdatalist = [] 
    for ee in energies:
      for aa in angles:
        params.target = tt
        params.beam   = bb
        params.energy = ee
        params.angle  = aa
        params.impacts = impacts
        wrapper.go(params)

      # extract coefficients and find transition angles
      fitdata  = help.extract_PDE_coefficients(params, angles)
      tdata    = help.find_pattern_transitions(fitdata)
      tdatalist.append(tdata)

    # after each energy sweep, plot and save the phase diagram 
    help.plot_energy_angle_phase_diagram(tdatalist) 
    plt.savefig("%s-%s-phase-diagram.svg" % (ion_species, target_species)) 
\end{lstlisting}
\caption{A program for generating angle-energy phase diagrams over many ion/target
combination. \label{alg: phase-diagram-listing}}
\end{algorithm}

\subsection{Binary Materials}

The simulation and analysis of single-impact events on binary targets
is easily performed within the PyCraters framework. In Ref.~\cite{norris-etal-NIMB-2013},
the Crater Function approach was extended to binary compound targets,
and estimates were made of several coefficients within a theory describing
the irradiation of such targets \cite{bradley-shipman-PRL-2010,shipman-bradley-PRB-2011},
for the case of GaSb irradiated by Ar at 250 eV. However, during this
process, a notable discrepancy between simulations and experiments
was observed -- whereas experiments observe gradual buildup of excess
Ga on the irradiated target, the simulations of impacts on GaSb indicated
a slight preferential sputtering of Ga, which would lead to excess
Sb. It was hypothesized that this discrepancy might be due to the
effect of Gibbsian surface segregation, whereby the atom with the
lower surface free energy (in this case Sb) migrates to the surface
\cite{yu-sullivan-saied-SS-1996-gibbsian-segregation}, where it is
more easily sputtered due surface proximity. Because the in-situ composition
profile is unknown, the targets used in the initial studies were homogeneous,
and could not capture this effect.

The PyCraters framework is an ideal tool to explore this hypothesis,
and a code listing which performs the needed simulations is provided
in Algorithm~\ref{alg: segregation-comparison}. In this example,
the user must move beyond built-in functionality, and begin to provide
some customization to the default settings. For instance, we specify
a list of amorphous targets with varying stoichiometries of the form
$\mathrm{Ga}_{1-x}\mathrm{Sb}_{x}$, and a simple user-supplied function
that allows a surface layer 1nm thick to be modified by enriching
it in Sb from $x$ to a plausible level of $y=2x-x^{2}$. After the
target is then irradiated at 1 keV by Ar+, and the relative sputter
yields of Ga and Sb (which appear in the zeroth moment of the crater
function) are extracted directly from the storage files, and a few
lines of code are written to plot these yields as functions of the
base Sb composition. 

\begin{figure}
\centering{}\includegraphics[width=3in]{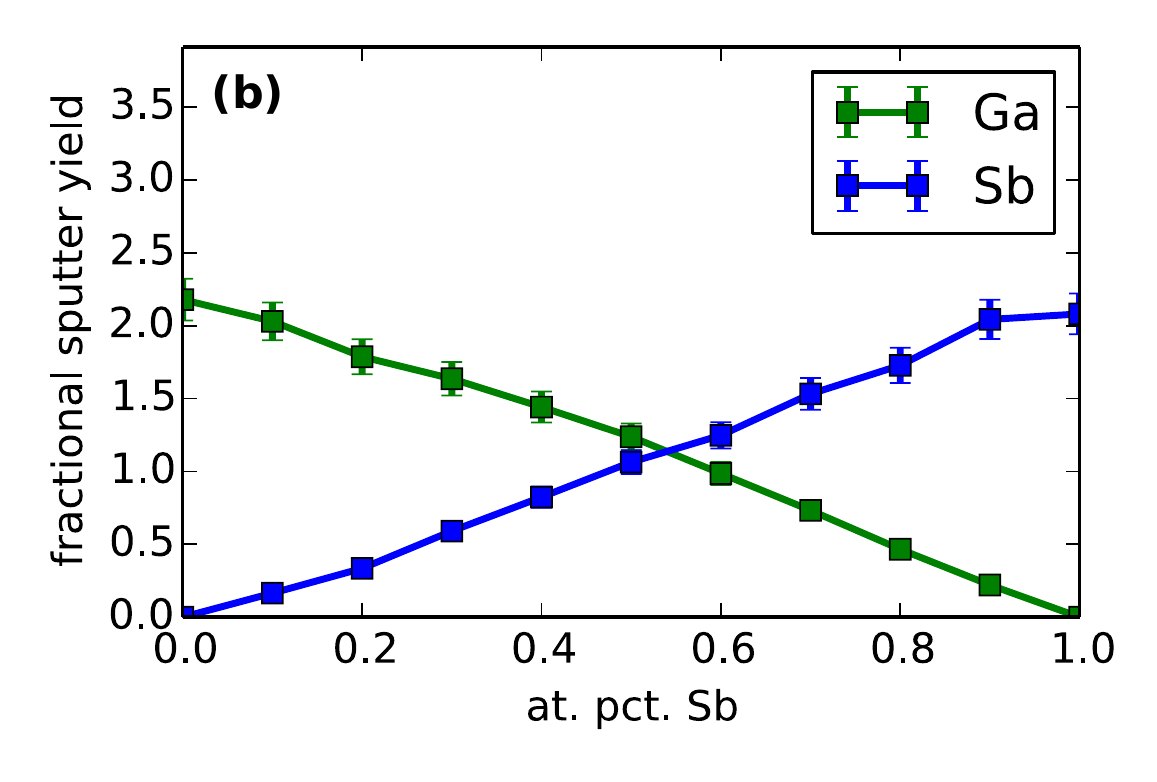}\includegraphics[width=3in]{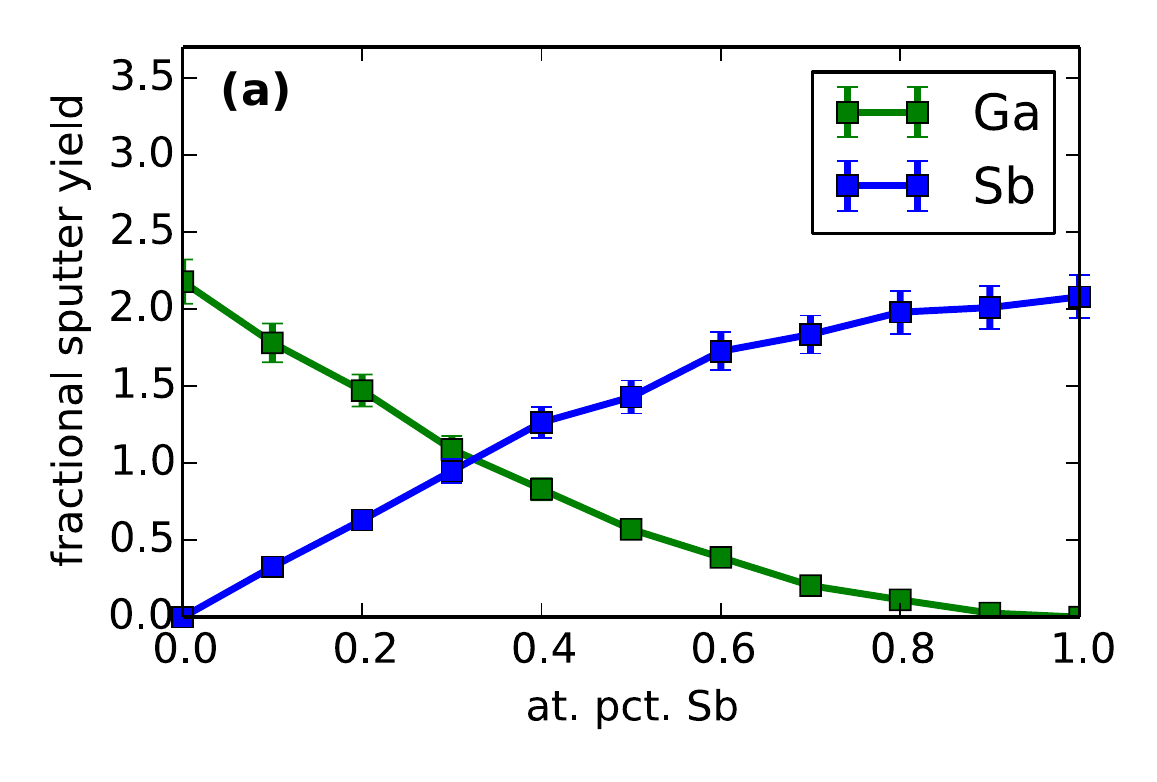}\caption{Estimates of the relative sputter yields for various amorphous compounds
$\mathrm{Ga}_{1-x}\mathrm{Sb}_{x}$ (a) if the target is homogeneous,
(b) if the target exhibits a 1nm layer enriched in Sb of the form
$\mathrm{Ga}_{1-y}\mathrm{Sb}_{y}$, with $y=2x-x^{2}$. The enrichment
of the surface layer produces a dramatic shift in the equilibrium
film concentration. \label{fig: relative-yields-with-and-without-segregation}}
\end{figure}

The results for both unmodified and modified targets are shown in
Figure~\ref{fig: relative-yields-with-and-without-segregation}.
Unsurprisingly, the modified target exhibits a greater sputter yield
of Sb than the unmodified target -- the top monolayer of this target
contains more Antimony, and most sputtering occurs from the top monolayer.
In fact, the new yields largely mirror the enriched layer composition.
Interestingly, however, this relatively small modification is entirely
sufficient to resolve the discrepancy between simulation and experiment
just described. Though the actual composition profile remains as yet
unknown, the results of this kind of experimentation with plausible
model targets suggests that this mechanism is likely sufficient to
explain the observed enrichment of Ga over time.

\begin{algorithm}
\begin{lstlisting}[basicstyle={\ttfamily},numbers=left]
# define a simple function of depth 
def concentration_function(params, depth): 
  cbase = [ a[1] for a in params.target ] 
  if depth > 10: 
    return cbase 
  else: 
    bSb = cbase[1] 
    lSb = 2.0*bSb - bSb**2 
    return [1-lSb, lSb]

# basic parameter setup 
params.target  = None 
params.beam    = "Ar" 
params.angle   = 0.0 
params.energy  = 1000
params.impacts = 1000
params.set_parameter("cfunc", concentration_function)

# set up targets at different concentrations
targets = [  ] 
for phi in np.linspace(0.0, 1.0, 11): 
  target = [["Ga", 1.0-phi],["Sb", phi]]
  targets.append(target)

# iterate over targets
for tt in targets: 
  params.target = tt
  wrapper.go(params)

# make plots
import libcraters.IO as io
m0e_avg = io.array_range( './', params, 'target', targets, 'm0e_avg' )
m0e_std = io.array_range( './', params, 'target', targets, 'm0e_std' ) 
Gaval = [ a[0] for a in m0e_avg ]
Sbval = [ a[1] for a in m0e_avg ]
Gaerr = [ b[0]/np.sqrt(params.impacts) * 1.97 for b in m0e_std ]
Sberr = [ b[1]/np.sqrt(params.impacts) * 1.97 for b in m0e_std ]

plt.figure(1, figsize=(4.5, 3.0))
pctSb_list = [ a[1][1] for a in targets ]
plt.errorbar(pctSb_list, -np.array(Gaval), yerr=Gaerr, label='Ga', fmt='gs-', linewidth=2)
plt.errorbar(pctSb_list, -np.array(Sbval), yerr=Sberr, label='Sb', fmt='bs-', linewidth=2)
plt.xlabel('at. pct. Sb')
plt.ylabel('fractional sputter yield')
plt.legend() plt.ylim(0, max(-np.array(Tvals))*1.7 )
plt.tight_layout() plt.show()
plt.savefig('relative-sputter-yields.svg')
\end{lstlisting}
\caption{Code to investigate the effect of segregation. \label{alg: segregation-comparison}}
\end{algorithm}

\subsection{Future work: Comparison of Methodologies}

An important future goal of the framework is to facilitate the comparison
of simulations performed using Molecular Dynamics to those using the
Binary Collision Approximation. For instance, using MD, Ref.~\cite{norris-etal-NCOMM-2011}
found redistributed atoms to contribute far more to the shape and
magnitude of coefficients in Eqs.~(\ref{eq: SXY-natcomm}) than did
sputtered atoms. However, using the BCA, Ref.~\cite{hofsass-APA-2014}
reports a reduced redistributive signal for the environment studied
in Ref.~\cite{norris-etal-NCOMM-2011}, and finds that sputtered
atoms are dominant for most angles at higher energies. The use of
MD vs BCA may be an important source of these conflicting results
-- in Ref.~\cite{bukonte-etal-NIMB-2013}, it was found that the
BCA reports significantly fewer displacements of small magnitude than
molecular dynamics. Because of the very large number of such displacements,
they may contribute significantly to the effect of mass redistribution,
and hence the BCA approach may systematically under-report the strength
of this effect. This question demands further study, and because the
extraction of the statistics and the estimation of coefficients are
generic processes independent of the solver within the PyCraters library,
it will be straightforward to use those common resources to identify
the implications of these observations for the specific statistics
required by crater function analysis.

In a related vein, several different procedures for extracting the
crater function $\Delta h$ and its moments $M^{\left(i\right)}$
from a list of atomic positions have been suggested in the literature
\cite{kalyanasundaram-etal-APL-2008,norris-etal-NCOMM-2011,hossain-etal-APL-2011,hofsass-APA-2014},
and the importance of the differences between the strategies has never
been investigated. The PyCraters library provides a natural environment
in which to conduct such studies. Because the framework provides the
results of simulations in a common format, variations on the statistical
extraction routine can be written in a way that is independent of
both the underlying solver \emph{and} the methods used to fit and
differentiate the results. This should allow easy comparison of the
different extraction methods for a variety of simulation environments,
and aid the identification of best practices for this important procedure.

\section{Summary}

In conclusion, we have introduced a Python framework designed to automate
the most common tasks associated with the extraction and upscaling
of the statistics of single-impact crater functions to inform coefficients
of continuum equations describing surface morphology evolution. The
framework has been designed to be compatible with a wide variety of
existing atomistic solvers, including Molecular Dynamics and Binary
Collision Approximation codes. However, in order to remain accessible
to first-time users, the details of each of these solvers is abstracted
behind a standardized interface, and much functionality can be accessed
via high-level functions and visualization routines. Although the
addition of much functionality is still in progress, the current codebase
is able to reproduce many important results from the recent literature,
and examples demonstrating these capabilities are included to facilitate
modification and additional exploration by the community. The project
is currently hosted on the BitBucket repository under a suitable open-source
license, and is available for immediate download.

\clearpage{}

\bibliographystyle{unsrt}
\bibliography{/home/snorris/Dropbox/literature/bibliography/tagged-bibliography}

\end{document}